\documentstyle[prd,preprint,aps]{revtex}

\begin{document}
\draft

\title{The Near-Linear Regime of Gravitational Waves in Numerical
Relativity }

\author{Peter Anninos${}^{(1)}$, Joan Mass\'o${}^{(1,2)}$, 
Edward Seidel${}^{(1,3)}$, Wai-Mo Suen${}^{(1,4,5)}$, and Malcolm 
Tobias${}^{(4)}$}

\address{${}^{(1)}$ National Center for Supercomputing Applications,
605 E. Springfield Ave., Champaign, Illinois 61820}

\address{${}^{(2)}$ Department of Physics,
Universitat de les Illes Balears, Palma de Mallorca E-07071, Spain}

\address{${}^{(3)}$ Department of Physics,
University of Illinois at Urbana-Champaign}

\address{${}^{(4)}$McDonnell Center for the Space Sciences,
Department of Physics,
Washington University, St. Louis, Missouri 63130}

\address{${}^{(4)}$Physics Department,
Chinese University of Hong Kong, Hong Kong}

\date{\today}

\maketitle

\begin{abstract}
\widetext
We report on a systematic study of the dynamics of gravitational waves
in full 3D numerical relativity.  We find that there exists an
interesting regime in the parameter space of the wave configurations:
a near-linear regime in which the amplitude of the wave is low enough
that one expects the geometric deviation from flat spacetime to be
negligible, but nevertheless where nonlinearities can excite unstable
modes of the Einstein evolution equations causing the metric functions
to evolve out of control. 
The implications of this for numerical
relativity are discussed.
\end{abstract}


\paragraph*{\bf Introduction.}
\label{introduction}

The study of gravitational waves and their interactions, i.e., the 
dynamics of spacetime in its pure vacuum form, is important for both 
theoretical and observational reasons.  However, due to its 
mathematical complexity and the need for previously unavailable large 
scale computational resources, apart from a few specialized solutions,
this area of research is largely unexplored. 
Previous analytic and numerical 
work on pure gravitational wave spacetimes, done in 1D or 2D, has led 
to many interesting results\cite{Allwaves,Yurtsever88}.
These discoveries raise interesting 
questions about waves in 3D.

Gravitational waves are also about to open up a fundamentally new area
of observation: gravitational wave astronomy.  A new
generation of detectors\cite{Abramovici92} should see waves for the
first time near the turn of the century.  Even though the observed
waves are expected to be weak here on Earth, they will have
been generated in regions with strong, highly dynamic and nonlinear
gravitational fields.  It is therefore essential to study
accurately both the strong and weak field regimes, as well as the long
term secular behavior in transitory intermediate regimes.

Thus we have been motivated to undertake a systematic study of
gravitational waves in 3D numerical relativity over the past few
years.  3D studies of general gravitational wave spacetimes are
different from lower dimensional studies in two very important
respects.  First, unlike in lower dimensional studies, there are no
assumed symmetries to fix the gauge.  One has to consider the full set
of gauge freedoms during the numerical evolution.  Second, 3D
calculations are bound to have lower resolution, and hence larger
errors.  This in turn makes it difficult to separate physical from
numerical and gauge effects.  These two effects are in fact related,
as finite differencing errors tend to excite all gauge freedoms
available.  The {\it excitation of unstable modes}, whether they
correspond to gauge freedoms or not, is the underlying principle of
the phenomena discussed in this paper.

We have
developed advanced 3D Cartesian numerical codes, using different numerical
techniques as well as different formulations of the Einstein equations,
to study gravitational waves.
Full details of our methods and codes
will be reported elsewhere\cite{Anninos94d}.
In this paper we focus on one important lesson we learned with the large
amount of numerical studies carried out using
these codes.

When the amplitude of the gravitational waves is low, the evolution is
linear.  At high amplitudes there are nonlinear
geometric effects: the energy-momentum carried by the waves
themselves changes the geometry of the spacetime so that the evolution
cannot be described by linear equations.  
We have seen both of these regimes in
our studies, as expected.  What was not anticipated is that
{\em beyond the linear regime, but before the amplitude is large
enough for nonlinear geometric effects to be significant, there
exists a regime where the evolution of waves, as described by the full
nonlinear Einstein equations, can differ dramatically from the
evolution described by the linearized versions of the Einstein
equations}.  This is what we
called the ``Near-Linear Regime'' (NLR).  The aim of this paper is to draw
attention to the existence of such a regime.

Many questions immediately come to mind: In what way do these effects
show up, how do they arise, and can they be controlled?  What are the
boundaries between the linear, near-linear, and nonlinear
regimes?  What are the implications in the numerical study of
gravitational waves, and gravitational systems in general?  Will such
effects be important for the numerical simulation of waves as weak as
those we expect to observe astrophysically?  These are the questions
we address in this paper.

\paragraph*{\bf Statement of the problem.}
\label{statement}
The evolution of the metric functions
is described by the Einstein equations in the standard 3+1 form
\begin{eqnarray}
\partial_t g_{ij} &=& - 2 \alpha K_{ij}+\nabla_i \beta_j+\nabla_j
\beta_i
\nonumber \\
\partial_t K_{ij} &=& -\nabla_i \nabla_j \alpha + \alpha \left(
R_{ij}+K\ K_{ij} -2 K_{im} K^m_j \right)  
\label{equations} \\ &\ & + \beta^m
\nabla_m K_{ij}+K_{im} \nabla_j \beta^m+K_{mj} \nabla_i \beta^m~. \nonumber
\end{eqnarray}
The question is: When will this evolution be different from what is
described by the linearized evolution equations, namely, by taking
$g_{ij} = \delta_{ij} + \epsilon h_{ij}$,
and keeping only terms to first order in $\epsilon$ in
Eqs.(\ref{equations})?  More precisely, we ask when
the discrete versions of Eqs. (\ref{equations}) yield
results significantly different from the discrete versions of the
linearized equations, using the same finite differencing techniques.
We use the same initial data, which satisfies the initial constraint
equations, for both the nonlinear and linear evolutions.

The purpose of this paper is to demonstrate, with concrete examples,
that significant nonlinear effects appear before one might expect.
Without careful analysis,
these effects can be mistaken for true
geometric effects, e.g, the formation of a spacetime singularity.
These effects are due to unstable nonlinear modes admitted by
Eqs.~(\ref{equations}) (and/or their finite differenced
versions), excitable by waves even with quite low amplitudes.

\paragraph*{\bf Quadrupole waves.}
\label{quadrupole}

The first example is of quadrupole waves similar to those studied 
in\cite{Teukolsky82}, except that we use an imploding-exploding wave 
configuration which is time symmetric\cite{Eppley79}, and we take the 
metric given in \cite{Teukolsky82} as the conformal part of the metric 
function. We then determine the conformal factor using the York
decomposition to solve the hamiltonian 
constraint equation\cite{York79}.  
The time symmetry of this imploding-exploding 
data automatically guarantees that the momentum constraint is
also satisfied.

We study a case where the wave is initially centered at the origin,
with a peak amplitude of
$A \equiv g_{ij} - \delta_{ij} \sim
10^{-3}$, and a width of order unity (we use $c=G=1$
throughout this paper).  The evolution of this initial data is
predictable: As the amplitude of the wave is of the order $A \sim
10^{-3}$, and both space and time derivatives can bring in only
factors of order unity, the energy density (e.g. Landau-Lifshitz
pseudo energy density) must be of order $A^2\sim 10^{-6}$.  With this
``wave energy density'' concentrated in a spatial dimension of order
unity, the strength of the ``gravitational potential M/R'' (or
spacetime curvature), can at most be of order $10^{-6}$.  This is approximately
$10^{6}$ times too weak to produce a significant effect on the
propagation of the wave, e.g.  to ``bend'' the null characteristics of
the spacetime.  Hence we expect the linear wave to expand outward just
as in flat space.  The scattering of the wave by itself, and any tail
effects would affect the metric functions to at most of order
$A^3 \sim 10^{-9}$, which is beyond the level of accuracy in our
simulations.  Hence the wave train should propagate
outward basically in a linear fashion, except for a slight delay in
phase.  

This picture is confirmed in Fig.~\ref{fig:fig1}a where we
show the evolution of a typical 
metric function $g_{xx}$ (solid lines) 
along the $z$ axis from the initial time $t=0$ to $t=3$,
with zero shift and unit lapse.
This is to be compared to what is
obtained with the same initial data, but evolved with the
linearized evolution equations (dotted lines).  At early times, the two
waveforms are completely indistinguishable.

Next we look at the evolution at late times 
in Fig.~\ref{fig:fig1}b and zoom into the origin 
where we expect the spacetime to have returned to flat space. 
The nonlinear 
results (solid lines) are basically one everywhere except for a dip near the origin.  
This is clearly due to nonlinear effects, as there is absolutely no dipping in 
the linearized treatment (dotted lines).  There are similar dips in $g_{yy}$ and 
$g_{zz}$, while the off diagonal metric functions develop a shear-like
structure.  These ``residual'' features will keep growing in time until 
the code crashes.
This behavior is independent of grid resolution and has a characteristic
growth rate that increases in proportion to $A^2$.

What causes this result?  There are four possibilities: (i)
nonlinear spacetime geometric effects which are not
captured by the order of magnitude estimate given above;  (ii)
nonlinearity coupled with finite differencing errors;
(iii) nonlinearity coupled with gauge effects;  (iv) 
bugs in the nonlinear evolution code. In principle, any of these
possibilities can cause this feature.

After much investigation\cite{Anninos94d} that includes 
numerous code tests, convergence studies, comparisons with linear evolutions,
different numerical techniques, 
and even different formulations of the Einstein equations,
we conclude that, in this case, it is (iii).  The nonlinearity excites an
unstable gauge mode of the evolution equations.  {\em A posteriori},
the effect is simple to understand.  The potential well
created by the initial wave packet, although quite small in amplitude,
sets the coordinate lines into motion, drifting towards each other.
As the wave propagates outward, the potential well (spacetime 
curvature) reduces
to zero, but the coordinate lines, once set into motion, will keep
drifting towards the center.  Hence over time, the distance between,
say, constant $x$ coordinate lines decreases and a dip in $g_{xx}$
develops. 

At what point does this effect appear (the lower boundary of the
NLR)?  This depends on the amplitude of the wave
and the length of time it is evolved.  We have observed dipping
for waves with magnitude $A \sim 10^{-4}$ 
and spatial extension the same as in Fig.~\ref{fig:fig1}a,
after it is evolved to about $t=10$.

At what amplitude would nonlinear geometric effects become
significant (the upper boundary of the NLR) for the quadrupole waves
at late time?  In Fig.~\ref{fig:fig1}c we show the 
Riemann invariant $J$\cite{Kramer80} obtained using nonlinear and linear evolutions
at time $t=4$, for various initial wave amplitudes $A$'s.
We see that despite the dipping of the metric, the invariants are basically the
same up to $A\sim 0.05$. Beyond that, the linear and nonlinear
results rapidly diverge, indicating nonlinear
geometric effects. 

The existence of unstable modes in the nonlinear evolution
equations in the NLR
highlights the importance of controlling the motion of the
coordinates.  In Fig.~\ref{fig:fig1}d we show the evolution of
$g_{xx}$ at various times using the minimal distortion
shift\cite{Smarr78}
and unit lapse.  We see that there is no dipping.
Details of this calculation and other methods in controlling the coordinate
motion with both shift and lapse will be discussed elsewhere\cite{Anninos94d}.

\paragraph*{\bf Colliding Wave Packets.}
\label{plane}

Next we discuss the collision of two plane wave packets. In this case a
completely different type of unstable mode is excited in the numerical
evolution of Eqs.~(\ref{equations}).  The 
initial data is given by two nearly gaussian packets of the form shown 
by the solid line at $t=0$ in Fig.~\ref{fig:fig5} for the metric 
function $g_{xx}$.  The evolution is carried out with $\alpha=1$, 
zero shift, and grid spacing $\Delta x=0.05$.  
The wave packets, which are initially located at $z=\pm 3$, travel
towards each other and
collide at the center of the grid at
$t=3$.
In the wake of the waves, following the collision, we see that 
the metric function $g_{xx}$ develops an upward drift.  
Comparing $g_{xx}$ from the linear (dashed line) and nonlinear
evolutions at $t=9$, we see clearly that the drift is a nonlinear effect.

Again we face the four possibilities pointed out above.  In fact
it is known that when two plane symmetric waves collide, a curvature
singularity is generated due to the focusing effects of the waves
~\cite{Yurtsever88} even for arbitrarily weak waves.
However, it is easy to show, based on the colliding packet study in
~\cite{Yurtsever88}, that the singularity will develop at a time
$t \sim \lambda^2/[(2\pi A)^2 \sigma]$
after the collision, where $\lambda$ is the characteristic wavelength, 
$\sigma$ is the characteristic width of the packet, and $A$ is the 
characteristic amplitude of the packet.  For the case here, with 
$\lambda \sim 1$, $\sigma \sim 1$, and $A \sim 10^{-2}$, we expect the 
singularity to appear at $t \sim 250$, which is far beyond any 
evolutions shown here.  Of course this estimate does not rule out that 
this drift is a ``precursor'' of the singularity, nor the possibility 
of other geometric nonlinear effects.

After much investigation\cite{Anninos94d} we confirm that the drift is
an unstable mode in the nonlinear evolution
equations.~(\ref{equations}).  We note that
this is not in contradiction to the expectation that the Einstein
equations are stable for weak waves (weak perturbations of the flat
spacetime).  It is the constraint equations that rule out these
unstable modes.  In our free evolution code the constraint equations
are not enforced.  This allows the unstable modes to develop after
they are excited by the numerical errors in the evolution. 
The amount it is excited depends
on the details of the numerical scheme.  The instability causes the
metric components in the wake of the waves to drift upwards according
to $g_{ij} - \delta_{ij} \propto -\ln(Ct + 1)$.  The time scale $1/C$
of the instability depends strongly on the finite differencing scheme
used.  For the same run parameters used in Fig.~\ref{fig:fig5}, we
find that $C=-1.6\times 10^{-4}$ for a staggered leapfrog scheme, and
a very different $C=7.9\times10^{-7}$ for a MacCormack scheme.
We stress that the unstable mode exists on the level of the evolution 
equation and is excited in
both of these two standard schemes, although with very different rates.

\paragraph*{\bf Conclusion.}
\label{conclusion}

We have shown that there exists a ``near-linear regime'' (NLR) in 
the numerical study of waves in general relativity, in which the 
amplitude of the wave is weak enough so that nonlinear geometric 
effects are unimportant, but nevertheless unstable modes of the 
Einstein evolution equations can be excited by the nonlinearity.  The 
instabilities manifest themselves as secular changes in the metric 
functions, which eventually grow to a level that affects the evolution, 
even crashing the code.  There are various types of unstable modes.  
We have shown two examples of such modes \cite{note1}.
In the quadrupole wave case, the instability is 
caused by nonlinearity coupled with gauge freedom.  In the colliding 
plane wave packets case, the instability is caused by nonlinearity 
coupled with finite-differencing errors.  The former effect conserves
the constraint equations, the latter does not. In general,
effects will be cross coupled and show up in many different forms.  Further 
details of this analysis, as well as studies of other gravitational 
wave spacetimes will be given elsewhere.  Our studies have been based 
solely on free evolutions. 
Unstable modes will be different for constrained evolutions.

The aim of this paper is to draw attention to the existence of the 
NLR, so that the large amount of work leading to our keen awareness of these
NLR phenomena need not be duplicated by other research groups in the field.
We believe the instabilities in this 
regime are important in general for numerical relativity, with implications
not just for gravitational waves, but also for evolving 
other 3D gravitational systems.  The determining factor of whether 
nonlinear effects are important or not depends on both the 
strength of the gravitational potential and
on the length of time the general relativistic system is evolved.
To determine gravitational waveforms for astrophysical 
events, e.g., inspiral coalescences of compact binaries, which are 
expected to be one of the important observable sources of 
gravitational radiation, we would like to be able to evolve the 
system, compact objects and waves, to hundreds or even thousands of 
$M$, where $M$ is the mass of the system.  For the success of such 
long time scale simulations, a full understanding and control of the 
kind of unstable modes pointed out in this paper are indispensable.

This research is supported by NCSA, PSC, and NSF grants Nos.  
PHY94-04788, PHY94-07882, ASC95-03978 and ASC93-18152.

\bibliographystyle{prsty}


\begin{figure}

\caption{(a) The metric function $g_{xx}$ is plotted along the $z$ axis 
for a quadrupole  wave. 
The wave peaks at the origin at $t=0$, and disperses outward. 
By $t=3$ the region near the origin returns basically to flat space,
as expected.  The same initial data is evolved with the nonlinear evolution  
equations (solid lines), and the linear evolution equations (dotted lines).
(b) $g_{xx}$ is shown
at later times $t=4,6,8,10$ for both linear (dotted lines)
and nonlinear evolutions (solid lines).  Note the
development of a dip near the origin with the nonlinear evolution.
(c) The Riemann invariant $J$ at the origin, is plotted against the initial
amplitude of the wave $g_{zz}-1$, for both the nonlinear
(solid line) and linear evolutions (dotted line) at $t=4$.  At small
amplitudes the two J's are the same, despite the dipping of the metric in
the non-linear evolution.  For perturbations of order
5\% we start to see deviations, indicating the presence of nonlinear
geometric effects.
(d) An evolution of the same quadrupole data, but
now evolved with the minimal distortion shift and $\alpha=1$.  Note
that the shift is able to hold the coordinate lines from drifting in.
}
\label{fig:fig1}
\end{figure}

\begin{figure}

\caption{The metric function $g_{xx}$ is shown for colliding plane wave
packets. The packets are initially centered at $z=\pm 3$ at 
$t=0$ and are moving towards each other.  They collide at
the origin at $t=3$, then pass through each other.
The solid lines show the nonlinear evolution.  Note the upward drift
in the ``wake'' regions at $t=9$.
Such drifting is absent in the linear evolution (dashed line).}
\label{fig:fig5}
\end{figure}

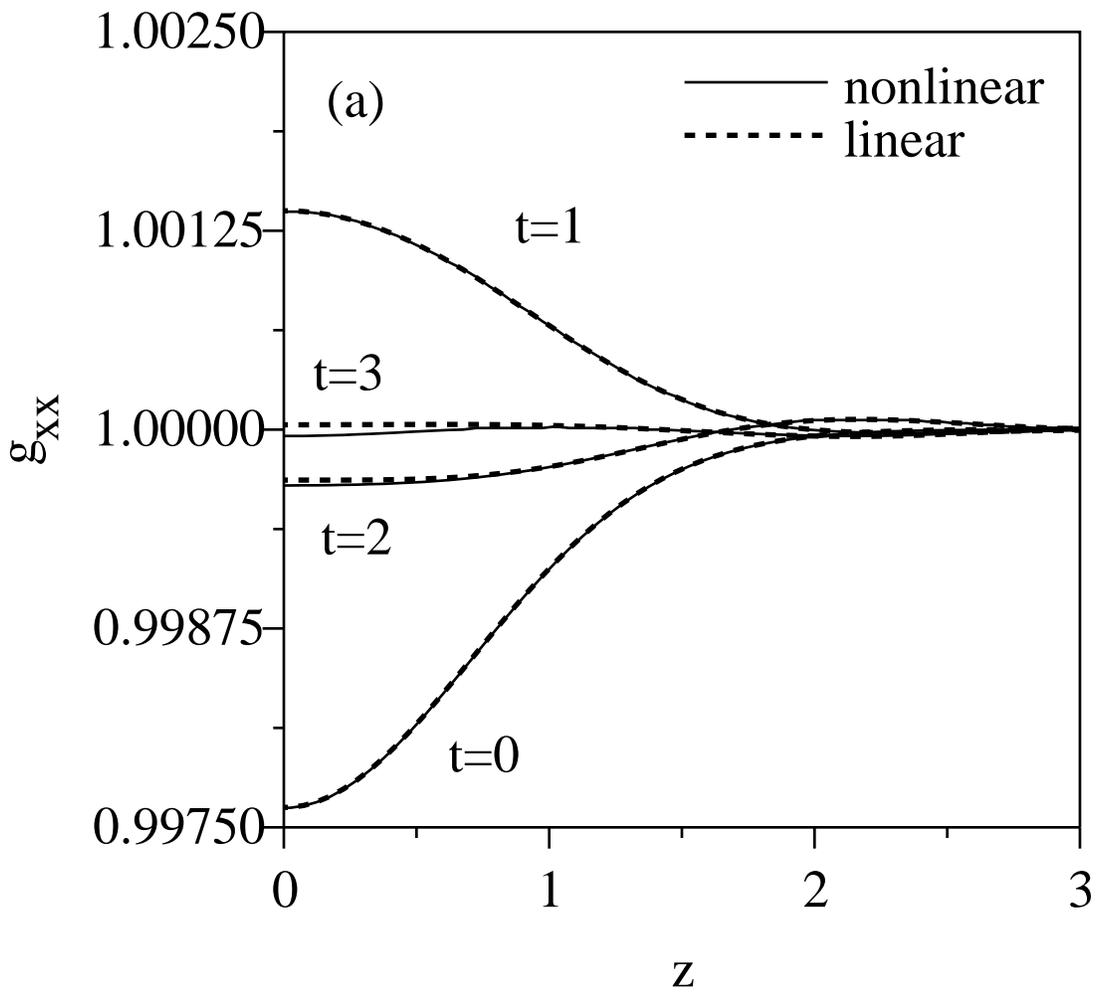
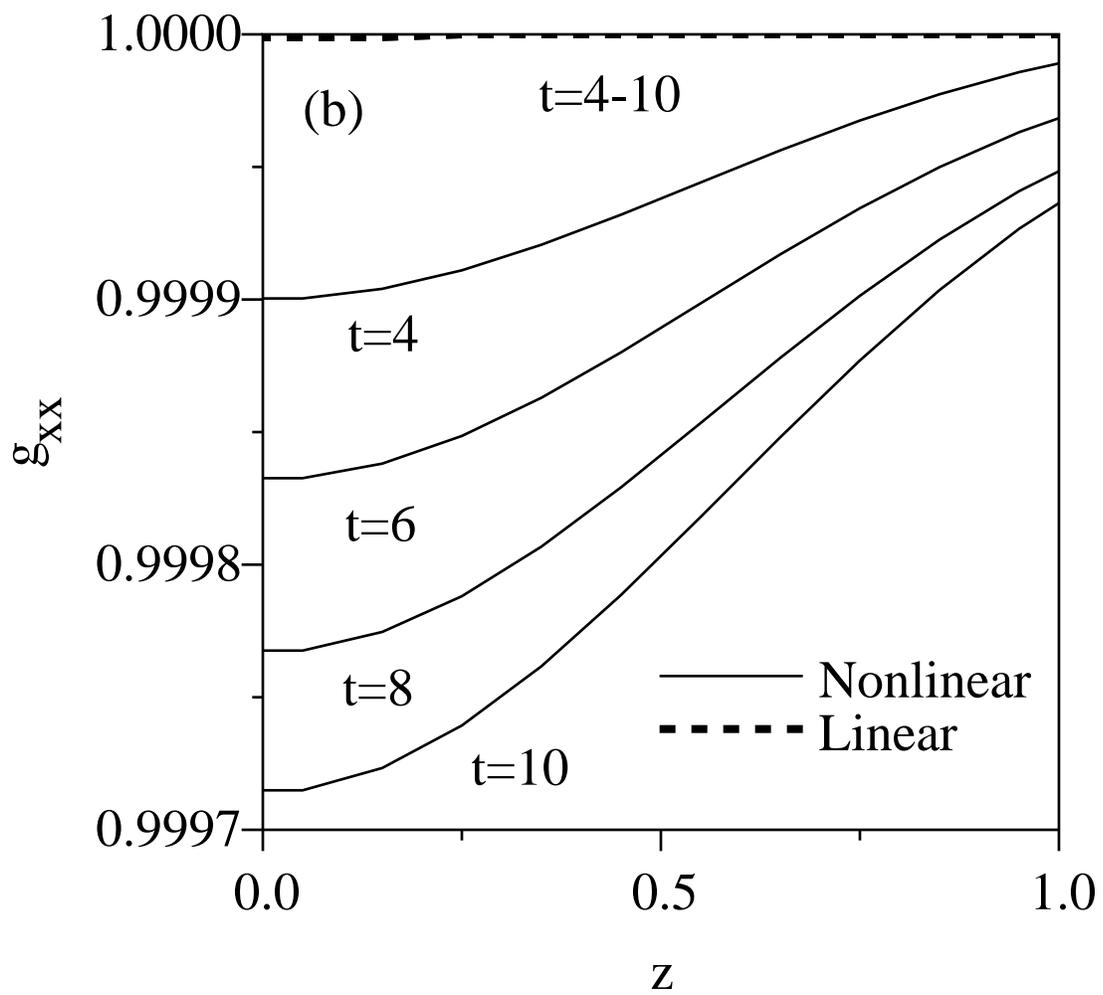
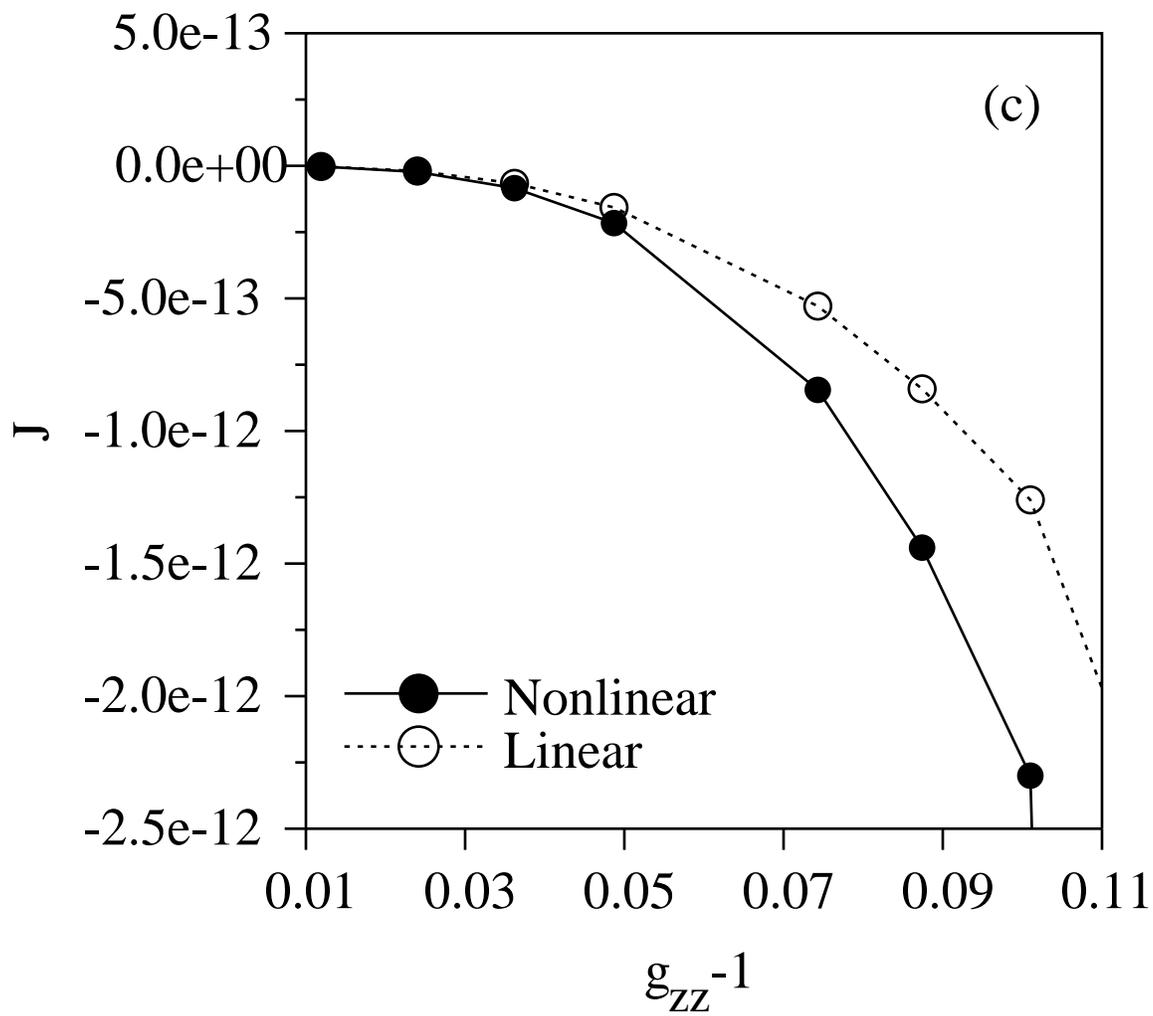
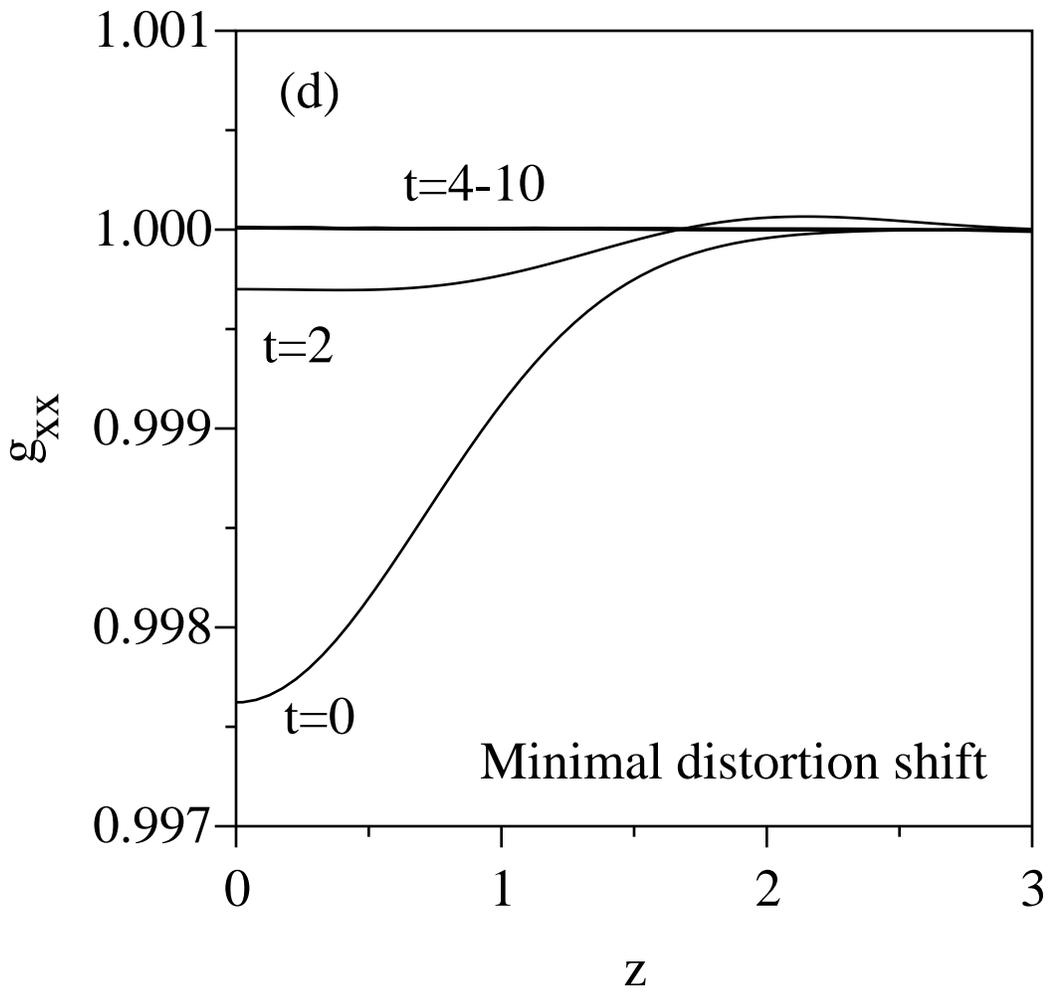
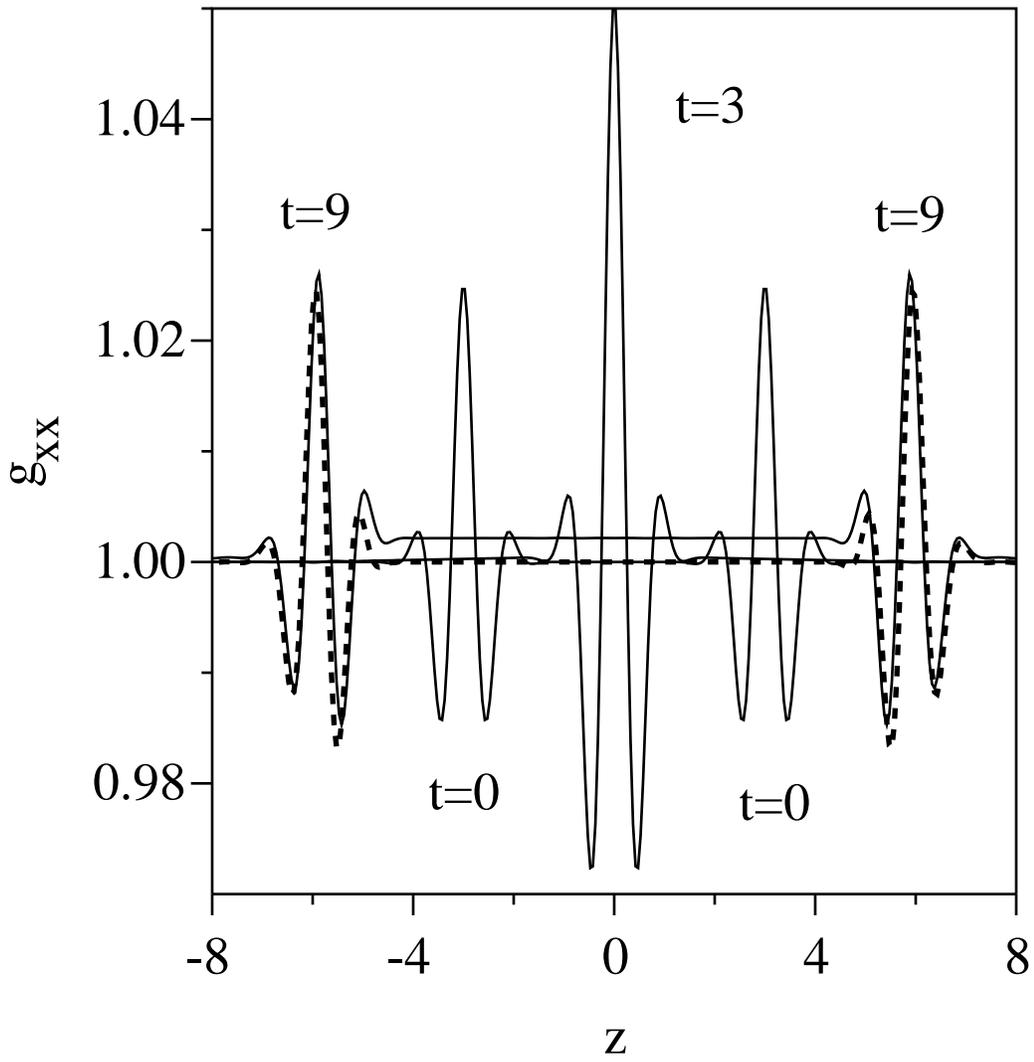
\end{document}